\documentclass[journal]{IEEEtran}
\IEEEoverridecommandlockouts
\normalsize

\ifCLASSINFOpdf

\else

\fi
\usepackage{soul}
\usepackage{array}
\usepackage{color}
\usepackage{amsfonts}
\usepackage{amssymb}
\usepackage{amsmath}
\usepackage[pdftex]{graphicx}
\usepackage{cite}
\usepackage{balance}
\usepackage{caption}
\usepackage{subcaption}
\usepackage{textcomp}
\usepackage{gensymb}
\usepackage{float}
\usepackage{comment}
\usepackage{subcaption}
\usepackage{tabularx}
\usepackage{multirow}
\usepackage{amsthm}
\usepackage{algorithmic}
\usepackage{algorithm}

\usepackage[utf8]{inputenc}
\begin{document}


\title{LLM-Integrated Digital Twins for Hierarchical Resource Allocation in 6G Networks}

\author{Majumder Haider, ~\IEEEmembership{Graduate Student Member,~IEEE}, Imtiaz Ahmed, ~\IEEEmembership{Senior Member,~IEEE}, Zoheb Hassan, ~\IEEEmembership{Member,~IEEE}, Kamrul Hasan, ~\IEEEmembership{Member,~IEEE}, and H. Vincent Poor, ~\IEEEmembership{Life Fellow,~IEEE,} 

}

\maketitle

\begin{abstract}
Next-generation (NextG) wireless networks are expected to require intelligent, scalable, and context-aware radio resource management (RRM) to support ultra-dense deployments, diverse service requirements, and dynamic network conditions. Digital twins (DTs) offer a powerful tool for network management by creating high-fidelity virtual replicas that model real-time network behavior, while large language models (LLMs) enhance decision-making through their advanced generalization and contextual reasoning capabilities. This article proposes LLM-driven DTs for network optimization (LLM-DTNet), a hierarchical framework that integrates multi-layer DT architectures with LLM-based orchestration to enable adaptive, real-time RRM in heterogeneous NextG networks. We present the fundamentals and design considerations of LLM-DTNet while discussing its effectiveness in proactive and situation-aware network management across terrestrial and non-terrestrial applications. Furthermore, we highlight key challenges, including scalable DT modeling, secure LLM-DT integration, energy-efficient implementations, and multimodal data processing, shaping future advancements in NextG intelligent wireless networks.
\end{abstract}


\vspace{-0.5cm}
\section{Introduction}

The next generation (NextG) cellular networks are expected to enable revolutionary applications such as holographic telepresence and multisensory extended reality (XR) while introducing unprecedented challenges to meet the performance requirements of such applications. With ultra-dense deployments, dynamic topologies, and diverse service requirements—ranging from terabit-speed ultra mobile broadband (uMBB) to mission-critical extremely-reliable low-latency communications (ERLLC)—NextG must seamlessly integrate space-air-ground infrastructures. These stringent demands significantly complicate radio resource management (RRM), necessitating intelligent, multi-layer resource allocation to optimize energy efficiency and maintain quality of experience (QoE) across heterogeneous networks \cite{shen2021holistic}. Achieving this requires a holistic, artificial intelligence (AI)-driven approach to network management that balances adaptability with computational efficiency.

Digital twin (DT) technology is a key enabler for tackling NextG networks by providing real-time virtual replicas of physical network elements to model, analyze, and predict network behavior accurately and insightfully. By synchronizing with real-time data, DTs facilitate proactive monitoring, predictive analytics, and closed-loop control, enabling NextG operators to anticipate network issues and optimize wireless resources before implementation in the live physical network. This capability is essential for managing the complexity of heterogeneous, AI-driven NextG networks, ensuring adaptive and efficient performance.

In contrast to other cyber-physical systems (CPSs), the rapidly changing nature of the wireless propagation environment makes it significantly challenging to replicate the physical surroundings precisely in the virtual domain. When integrated with advanced AI models, high-fidelity DT-generated datasets can enhance context-aware predictions and reduce this discrepancy. However, traditional task-specific AI models lack general decision-making capabilities, as they are trained on limited multi-modal datasets. Moreover, executing multiple AI models for different tasks imposes significant computational overhead and increases hardware complexity, necessitating a more scalable and adaptive AI-driven approach. 

Conventional task-specific DTs offer fine-grained models but lack holistic decision-making. They rely on domain-specific models and human-defined rules, making them insufficient for handling the multimodal data and cross-domain complexities of NextG networks. Large language models (LLMs) serve as a powerful complement to DTs in NextG networks. As generative AI (GenAI) systems, LLMs are trained on diverse text data, enabling them to generalize across multi-modal environments and extract insights from numerical telemetry to unstructured logs \cite{bariah2024large, zhang2024large, lee2024llm}. By integrating LLMs, DTs transition from descriptive virtual models to cognitive network agents, capable of detecting patterns, identifying anomalies, and generating optimized policies in natural language. This LLM-DT synergy facilitates holistic network management, merging technical efficiency with semantic-level reasoning, a necessity for the complex, multimodal nature of NextG communication networks.
A summary of some recent papers [2]-[7] pertinent to the LLMs in the context of wireless networks is highlighted in Table I. 
\begin{table*}[t!]
\centering
\captionsetup{justification=centering}
\caption{Refined LLMs in the context of wireless networks}
\begin{tabular}{|p{0.99cm}|p{3.5cm}|p{3.5cm}|p{3.5cm}|p{4.0cm}|}
\hline
\bfseries{Paper} & \bfseries{Focus Area} & \bfseries{Contextual Learning} & \bfseries{Use Case} & \bfseries{Performance Accuracy}  \\
\hline
\cite{bariah2024large} & Represents visions of LLMs for the domain of wireless networks  & Agent and RL based  & Mentioned the potential of LLMs in several use cases, but implementation is not shown   &  No specific results are shown  \\
\hline
\cite{zhang2024large} & Refines LLMs for intrusion detection  & Contextual text data using real network dataset  & Cybersecurity: intrusion detection   & LLMs outperform CNNs   \\
\hline
\cite{lee2024llm} & Considers LLMs for radio resource allocation  & Synthetic contextual text dataset  & Resource allocation: wireless transmit power prediction & Compared different LLMs' performance to determine energy and spectral efficiency   \\
\hline
\cite{jiang2024large} & LLMs to meet the anticipated demands of 6G communications & Multi-agent based  & Semantic communications &  Performance comparison pertinent to LLMs is not shown  \\
\hline
\cite{xu2024large} & LLMs to meet the anticipated demands of 6G communications  & Multi-agent and model caching based  & Vehicular accident report generation & No comparative performance results are shown   \\
\hline
\cite{hu2024self} & LLMs for wireless traffic prediction  & Feedback based learning refinement using real dataset  & Wireless traffic prediction & LSTM RNNs outperform LLMs    \\
\hline

\end{tabular}
\label{tab2}
\end{table*}
Among others, a notable contribution has been made in \cite{huang2024digital}, which proposed GenAI and DT-integrated architecture for network analytics and management.

However, this article emphasizes LLM-driven DTs for network optimization (LLM-DTNet), a novel framework that integrates virtual protocol stack layers with LLMs to enhance NextG wireless resource management. LLM-DTNet features a multi-layer protocol structure augmented with AI-driven analytics and a high-level LLM-based decision module. At the lower layer, distributed micro DTs—virtual replicas of network segments (e.g., cells, edges, or slices)—continuously process telemetry and contextual data. These models leverage embedded AI algorithms for traffic prediction, interference mapping, and real-time network state analysis. At the top of this hierarchy, the LLM-engine orchestrator synthesizes multimodal and wide-area data, including numerical metrics, predictive insights, and text-based policies, to generate coherent and adaptive RRM decisions. 

A centralized controller is inadequate for managing NextG's dynamic RRM demands in real time. In the proposed LLM-DTNet framework, lower-layer DT instances (e.g., per cell, base station, or service) adapt to rapid radio/link state variations and execute localized resource adjustments. The upper-layer LLM agent integrates insights from these distributed twins, optimizing global objectives, inter-cell coordination, and cross-domain trade-offs. This division enables real-time control at lower layers while ensuring strategic, network-wide optimization at higher layers. The contributions of this paper are as follows:
\begin{itemize}
\item We propose LLM-DTNet, a hierarchical framework integrating DTs and LLMs for intelligent, context-aware RRM in NextG networks, enabling adaptive and scalable optimization.
\item We discuss the fundamentals and design considerations of LLM-DTNet, covering its multi-layer DT architecture, AI-driven analytics, LLM-based decision-making, and deployment strategies across centralized, semi-centralized, and decentralized implementations.
\item We demonstrate its applicability and outline future directions, showcasing LLM-DTNet's role in terrestrial and non-terrestrial networks, while addressing real-time data synchronization, multimodal processing, security, energy efficiency, and scalable DT modeling for NextG networks.
\end{itemize}

\section{Fundamentals of LLM-DTNet}\label{sec2}
This section describes the architecture of the proposed LLM-driven and AI-engine-assisted RRM framework and its functional elements. Moreover, while describing the functions, we highlight the potential configurations of the proposed framework.
\subsection{Architecture of the Proposed Framework}
\begin{figure*}[h!]
\centering
\includegraphics[width = 16cm, height = 10cm]{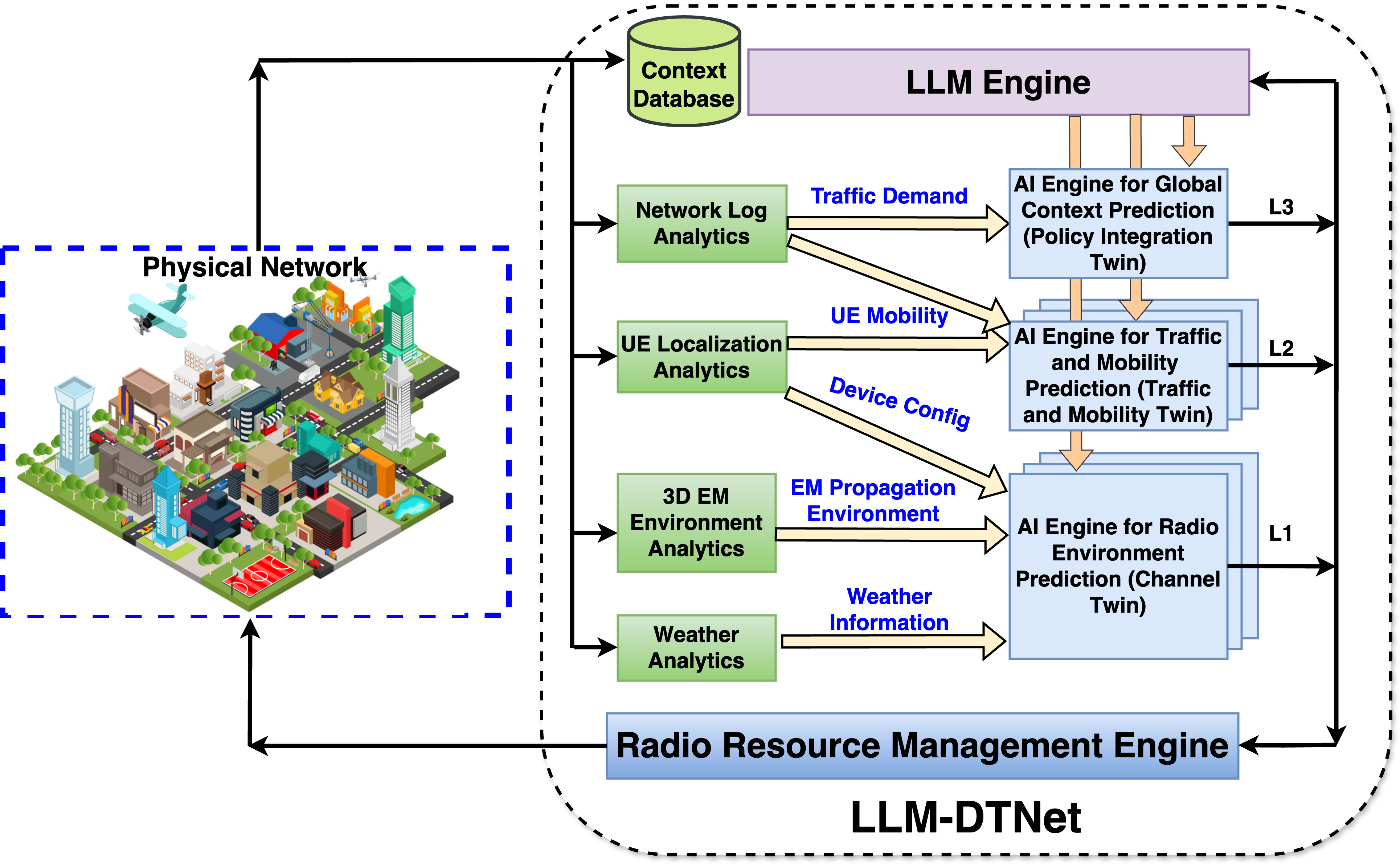}
\captionsetup{justification=centering}
\caption{Architecture of the proposed LLM-DTNet framework.}
\label{fig1}
\end{figure*}

\noindent Fig. \ref{fig1} shows the architecture of the proposed LLM-DTNet framework for multi-layer and contextual resource allocation for NextG wireless networks. In this architecture, the analytics modules, along with their task-specific AI engines, the context database, and LLM-based decision-making components constitute LLM-DTNet and synergistically operate to allocate and manage optimal resources for the physical network via virtual-to-physical (V2P) and physical-to-virtual (P2V) interfaces. 

\noindent\textbf{Physical Network:} Depending on the service requirements, the physical network includes network entities--such as base stations (BSs) / access points (APs), user equipment (UE), and emerging components (e.g., unmanned aerial vehicles (UAVs),  intelligent reflecting surfaces (IRSs), etc.)—along with dynamic wireless propagation conditions. These entities generate real-time data on user mobility, channel quality, and device capabilities. 

\noindent\textbf{Context Database:} All relevant data—ranging from real-time measurements (e.g., channel states, radio link quality, traffic logs) to metadata (e.g., user quality-of-service (QoS) profiles, device capabilities)—is aggregated and stored in the context database. This database serves as a unified repository, ensuring multi-modal data availability for cross-layer prediction and control tasks.

\noindent\textbf{Layer-Specific Analytics with AI Engines:} A high-fidelity DT environment mirrors the physical world by accurately modeling its RF propagation characteristics and diverse network elements (for instance, reflecting building layouts, material electromagnetic properties, or traffic distributions). This modeling is achieved by a combination of AI-assisted radio emulation such as Ray tracing, enabling highly granular representations of real-world conditions. Specialized AI-driven predictors refine the emulated context by compensating for any modeling gaps and extracting feature-rich insights. These task-specific AI models focus on lower-level radio resource metrics (e.g., channel gains, interference, modulation schemes) or mid-layer network analytics (e.g., traffic loads, session continuity), ensuring near-real-time accuracy in the DT’s representation of physical processes.

\noindent\textbf{LLM Engine:} The LLM engine is the upper layer manager of LLM-DTNet. In particular, this engine has two functionalities: a) twin management for L1, L2, and L3 engines and b) context analytics for the RRM engine. As a twin management entity, the LLM engine monitors the life cycle of layer engines and recommends when and how to update. The LLM engine captures three types of data to produce context analytics reports. This report contains a set of important contexts that are relevant to specific scenarios. This functional block provides the report to the RRM engine and fine-tunes the RRM algorithm. High-level spatial and temporal situation-awareness data include the type of applications (video, URLLC, IoT), the operating scenario (industrial, indoor, outdoor, rural, urban, semi-urban), possible coexistence settings (e.g., Wi-Fi–5G, non terrestrial network (NTN)–terrestrial network), the operating frequency bands (FR3, FR2), and any relevant regulatory policies. Contextual data encompasses user mobility patterns, traffic demand, user density, weather conditions, terrain features, building geometries (3D), and foliage. Network data incorporates performance indicators such as signal-to-interference plus noise ratio (SINR) and coverage maps, along with multipath channel profiles (e.g., delay, amplitude, angles of arrival and departure) and spectrograms for comprehensive channel analytics.

\noindent\textbf{RRM Engine for Resource Control:} The RRM engine optimizes the resources based on the information provided by L1, L2, L3, and LLM engines. Integrating a vast amount of data for advanced RRM is challenging, as it requires significant data collection. LLMs provide a feasible approach to this problem while combining the unique capability of both LLM and L1/L2/L3 engines. Note that not every context is important for all scenarios, and DT does not need to emulate all contexts for every scenario. For example, weather impact is crucial for FR3/FR2 bands but has minimal impact on FR1 band. Similarly, dynamic variation of multi-path channel gain is crucial for vehicle networks but may be less important for Fixed Wireless networks. Thus, the selection of appropriate contexts with the assistance of the LLM engine is crucial for real-time network management and orchestration.

\subsection{Implementations of the Proposed Framework}
\noindent\textbf{Configurations of LLM-DTNet:} Depending on the target use case, the proposed framework adopts one of the three possible configurations: fully-centralized hierarchical (FCH), semi-centralized hierarchical (SCH), and decentralized hierarchical (DCH). In the FCH model, both the lower and mid-layer modules and the RRM optimization engine reside in the cloud, enabling a global, system-wide view at the expense of higher latency. By contrast, SCH distributes L1/L2/L3 DT functionalities across edge nodes while preserving a cloud-based global RRM. The SCH architecture can better accommodate moderate latency demands while retaining some benefits of the centralized control. Finally, DCH implements fully disaggregated low- and mid-layer DT and RRM components across the network to deliver near real-time responsiveness for applications such as adaptive beamforming, interference management, etc. In practice, large-scale scenarios with strict time sensitivity favor the DCH paradigm, whereas less delay-sensitive tasks (e.g., spectrum assignment, network fault detection, etc.) align naturally with FCH or SCH. These architecture choices can be encapsulated in game-theoretic terms: the decision-making process in FCH maps to a single-leader single-follower Stackelberg model, SCH employs a single-leader multi-follower approach, and DCH is represented by a multi-leader multi-follower paradigm. This game-theoretic abstraction not only captures tradeoffs in computational overhead and implementation complexity but also provides a principled way to design and analyze distributed versus centralized strategies under diverse NextG requirements.  

\noindent\textbf{Physical-to-Virtual and Virtual-to-Physical Interfaces:}
Communication between the physical network and its DT counterpart (i.e., P2V and V2P) can be realized over diverse connectivity technologies depending on latency, bandwidth, and deployment constraints. For example, FCH typically leverages high-capacity optical fiber links from BSs/APs to the cloud, whereas SCH/DCH arrangements can rely on mmWave or FR2 backhaul for edge-based DT modules. In some environments, low-frequency RF links (e.g., sub-6 GHz) or line-of-sight microwave can facilitate robust data exchange in remote or hard-to-wire locations. This flexibility ensures that measurement data (physical-to-virtual) and RRM directives (virtual-to-physical) flow seamlessly despite heterogeneous deployment conditions, enabling adaptive and near-real-time coordination across the entire LLM-DTNet framework.

\section{Design Considerations of LLM-DTNet}
In this section, we discuss the significance of designing the hierarchical architecture of LLM-DTNet, the role of the task-specific AI engine for each layer, and the contribution of the LLM engine to overseeing RRM and AI engines.

\subsection{Hierarchical Architecture of LLM-DTNet}
In the proposed LLM-DTNet framework, each network layer (L1, L2, and L3) features a dedicated analytics block and a task-specific AI engine, ensuring that the DT accurately captures its corresponding portion of the physical environment. The ultimate goal is to bridge the gap between raw simulation outputs (e.g., from Ray tracing or synthetic traffic models) and the real-world measurements that reflect actual propagation and demand patterns. By refining simulated data through carefully trained AI models, each layer’s twin attains high fidelity while remaining computationally viable.

\noindent\textbf{Layer 1 (L1) — Realistic Radio Environment Modeling:} \textit{\underline{Analytics Block:}} The L1 analytics block manages low-level parameters such as instantaneous channel gains, interference profiles, and multipath conditions. It uses a geometric or Ray tracing engine (e.g., SIONNA) to generate a baseline set of channel parameters, informed by 3D models of terrain, buildings, and environmental factors.
\textit{\underline{AI Engine (Radio Environment Prediction):}} Because pure Ray tracing often overlooks subtle real-world effects (e.g., foliage scattering, unmodeled reflectors), the L1 AI engine correlates measurement data—collected from on-site drive tests or user equipment (UE) feedback—with the Ray tracing outputs. This correlation is learned via supervised or semi-supervised learning, producing an adjustment factor to correct the synthetic channel gains \cite{haider2025digital}. For instance, if a channel measurement reveals stronger diffraction paths than predicted, the AI engine updates the model’s attenuation coefficients accordingly. The net result is an enhanced channel twin that yields realistic PHY-layer metrics (e.g., SINR), enabling accurate decisions on beamforming, multi-antenna precoding, and link adaptation \cite{haider2025digital}. 

\noindent\textbf{Layer 2 (L2) — Network-Wide Traffic and Mobility Analytics:} \textit{\underline{Analytics Block:}} The L2 analytics block focuses on aggregated or cluster-level insights, such as traffic load across multiple cells, user mobility patterns, and session-level QoS data. It may start with coarse, statistical models (e.g., Markov chain mobility or average throughput patterns) derived from partial logs, which typically lack context on micro-level variations in dynamic networks.
\textit{\underline{AI Engine (Traffic and Mobility Prediction):}} To reconcile these statistical models with reality, the L2 AI engine fuses additional measurements—such as user location updates, inter-cell handover logs, backhaul utilization data, and spatiotemporal traffic bursts. By applying data-driven approaches (e.g., LSTMs for mobility forecasts or generative models for traffic surges), this engine refines the raw L2 twin, aligning it with ongoing real-world patterns. For example, if a new hotspot emerges around a stadium on game day, the AI engine quickly updates the traffic matrix so that the L2 twin mirrors this sudden load variation. This ensures that decisions on load balancing, interference coordination, or slice admission at a regional level closely match actual network demands.

\noindent\textbf{Layer 3 (L3) — Global Context and Policy Integration:} \textit{\underline{Analytics Block:}} At L3, the analytics block produces a system-wide perspective, stitching together key performance indicators (KPIs) from L2 blocks (e.g., total throughput, coverage distribution) and encompassing contextual information (e.g., regulatory constraints on FR2/FR3, upcoming large-scale events, or cross-domain interactions with non-terrestrial networks). Initially, it might rely on purely top-down planning frameworks or historical operational metrics for large-scale decisions.
\textit{\underline{AI Engine (Holistic Network Management):}} The L3 AI engine refines this global twin by reconciling high-level forecasts with ground-truth operational results—such as network-wide user changes, macro QoS trends, or previously unmodeled constraints (e.g., sudden changes in spectrum policy). A knowledge-based or reinforcement learning (RL) approach can be used here, which continuously learns from L2 feedback about actual network performance. Over time, this AI engine adjusts system-level policies (for example, deciding how to allocate network slices, when to trigger large-scale changes in resource partitioning, or how to accommodate new service classes like URLLC). By aligning top-down planning with real performance outcomes, the L3 twin stays agile in a dynamic NextG environment.

\noindent\textbf{Coordination Across Layers:} Although each layer operates with its own analytics/AI block, the layers are tightly coupled. \textit{\underline{Bottom-Up Corrections:}} L1’s precise radio environment refinements feed into L2’s traffic and mobility predictions, ensuring that mid-layer forecasts do not assume overly idealized channel conditions.
\textit{\underline{Top-Down Guidance:}} L3’s policies (e.g., updated slicing rules or QoS targets) inform L2’s resource constraints (e.g., maximum permissible load per cell cluster), which in turn adjust L1’s scheduling and link-adaptation algorithms.

\subsection{Design of LLM Engine}
The LLM engine serves as the central orchestrator of the proposed LLM-DTNet, harmonizing the outputs from each layer's AI engine (L1–L3) with attaining network objectives (e.g., throughput, latency, energy efficiency) leveraging the attention mechanism of the transformer model \cite{vaswani2017attention}. Designing this engine in a manner that effectively addresses the complexities of RRM requires balancing domain specialization and computational feasibility, while handling real-time constraints. Below, we outline a principled design strategy for the LLM engine, focusing on its training paradigm, operational modes, and interaction with lower-layer twins. 

\noindent\textbf{Pre-Trained vs. Domain-Retrained LLM:} \textit{\underline{Hybrid Training Paradigm:}} An off-the-shelf, large-scale LLM (e.g., GPT or Llama variants) provides a strong baseline for language reasoning and general knowledge representation. However, default pre-training on internet-scale text rarely includes the specialized vocabulary and numeric relationships central to wireless communications. To address this gap, a domain-adaptation phase is essential: we leverage Retrieval-Augmented Generation (RAG) or fine-tuning with curated datasets (e.g., 3GPP standards, wireless propagation models, traffic logs) to train the LLM about the specific structure and constraints of NextG networks \cite{10599304}.
\textit{\underline{Efficient Fine-Tuning:}} Full-scale re-training of billions of parameters is seldom practical in time-sensitive, resource-constrained environments. Instead, we may use parameter-efficient approaches such as LoRA (Low-Rank Adaptation) or adapters to add domain-specific knowledge without retraining the entire model. This results in an LLM that understands standard radio resource constructs (e.g., “beam index,” “carrier aggregation,” “path loss exponent”) and can contextually map them to the local scenario.

\noindent\textbf{Context-Aware Prompting and Knowledge Management:} \textit{\underline{RAG for Real-Time Context Integration:}} Even after domain adaptation, the LLM benefits from continuous updates on the network’s state, e.g., new measurement logs or ephemeral policy changes. By adopting a RAG pipeline, the system retrieves up-to-date network metrics and scenario details (e.g., FR2 path loss under bad weather) from a dedicated knowledge base or from the context database. This retrieval step ensures that the LLM always grounds its decisions in the latest network status, without requiring repeated large-scale fine-tuning.
\textit{\underline{Dynamic Prompt Construction:}} The LLM engine uses carefully designed prompts to capture relevant predictions from the specialized AI engines. For instance, a prompt might include: ``Traffic Twin indicates a +20\% load spike in Cell \#4,'' ``Channel Twin warns of strong interference from new UAV relays,'' ``Mobility Twin reports high user velocities.'' Based on these items, the LLM can propose resource allocations (e.g., ``increase scheduling priority for Cell \#4, reduce maximum MCS near UAV coverage'') or request further clarification from a particular AI engine.

\noindent\textbf{Role in Oversight and Strategy Synthesis:} \textit{\underline{Policy Refinement and Conflict Resolution:}} Each layer’s AI engine offers localized optimization insights, such as channel gain corrections at L1 or traffic load predictions at L2. However, these micro-level objectives can conflict: boosting throughput in one cell may degrade performance elsewhere. The LLM engine, enriched with domain knowledge, reconciles such conflicts by comparing predicted outcomes from multiple layers. It can, for instance, weigh the relative benefits of lower-latency scheduling (proposed by the Mobility Twin) against a system-wide interference penalty (revealed by the Channel Twin).
\textit{\underline{Real-Time RRM Decision Drafting:}} Once the LLM engine synthesizes these insights, it crafts a set of recommended allocation actions (e.g., ``reassign 20 MHz from Cell \#2 to Cell \#4, reduce power in overlapping beams''). These are then passed to the RRM engine, which executes them in near real time, possibly after a short confirmatory feedback loop from L1–L3. As such, the LLM effectively translates high-level problem statements (e.g., meet URLLC targets for an autonomous vehicle route) into feasible, context-aware solutions.

\noindent\textbf{Adaptive Updates to AI Engines:} \textit{\underline{Monitoring Model Performance:}} The LLM engine can track how well each AI engine’s outputs match subsequent real measurements. For instance, if the Channel Twin consistently underestimates path loss in an urban canyon, the LLM can flag this and trigger a re-training or hyperparameter adjustment in the L1 AI engine.
\textit{\underline{Selective Re-Training or LoRA Injection:}} By analyzing performance logs, the LLM engine pinpoints the domain areas (e.g., new frequency bands, novel mobility conditions) that require refined or expanded training data. It then recommends selective re-training or the injection of new LoRA modules tailored to those conditions. This approach localizes the computational burden to the modules in need of improvement rather than blindly retraining the entire system.

\subsection{Design Considerations for the RRM Engine}
The RRM engine serves as the central decision-maker in LLM-DTNet, tasked with transforming the refined context from each layer’s AI engine (L1–L3) and the LLM engine into actionable, system-wide allocation policies. 

\noindent\textbf{Formulation of the Optimization Problem:} \textit{\underline{Multi-Layer Data Aggregation:}} The RRM engine integrates lower-layer analytics (e.g., channel quality, spectral efficiency) from L1, cluster-level insights (e.g., load balances, mobility trajectories) from L2, and top-level strategies (e.g., slicing constraints, QoS priorities) from L3. Additionally, the LLM engine provides context reports (e.g., weather or deployment-specific constraints) gleaned via prompt-based reasoning. Consequently, the RRM engine can formulate holistic utility functions or constraints incorporating short-term link states alongside broader network objectives.
\textit{\underline{Complexity vs. Real-Time Constraints:}} Depending on the network scenario (e.g., URLLC with millisecond-level delay sensitivity vs. broadband best-effort), the RRM problem can range from small-scale convex programs to large-scale, NP-hard formulations. Designers must choose an appropriate solution scheme—analytical optimization, metaheuristics, or data-driven models—based on feasibility within real-time constraints. The merits and demerits of design considerations for the RRM engine are further illustrated in Table II.

\begin{table}[h!]

\centering
\captionsetup{justification=centering}
\caption{Design for Intelligent RRM Engine}
\begin{tabular}{|p{1.3cm}|p{1.3cm}|p{1.3cm}|p{1.4cm}|p{1.4cm}|}
\hline
\bfseries{Solution Method} & \bfseries{Example} & \bfseries{Advantage} & \bfseries{Disadvantage} & \bfseries{Use Case}  \\
\hline
Conventional classical optimization  & Branch-and-bound, Lagrangian duality, or Benders decomposition & tractable solution for well-defined convex or near-convex optimization problem formulations  & high computational complexity and overhead  &  transmit power allocation  \\
\hline
Neural network optimization & DNN, DRL, GenAI & intractable solution for complex optimization problems with low complexity and overhead  & require large datasets for learning   & especially suitable for time-sensitive applications like dynamic beam steering.   \\
\hline
LLM-assisted optimization & LLAMA, GPT, BERT & Excellent generalization ability, heuristic refinements  & precise contextual learning, require large computational resources for training  & situation-aware cell user association and traffic management.   \\
\hline
\end{tabular}
\label{tab2}
\end{table}

\section{Use Cases of LLM-DTNet in Next-G Wireless}
The objectives of NextG wireless networks are expected to shift from being just focused on communications to being service-oriented. Orchestration in managing different layers of data to make prompt decisions to meet dynamic user demands in such complex wireless networks is vital. This section highlights the potential use cases of LLM-DTNet in hierarchical network management in both terrestrial and non-terrestrial networks.

\subsection{Use Case: Terrestrial Networks}

\subsubsection{Proactive Resource Management}
Thanks to the multi-modal context availability and what-if analysis, LLM-DTNet can enable proactive network management in industrial environments by using real-time telemetry and extended reality interfaces to predict network performance and optimize configurations \cite{sudhakaran2024wireless}. For instance, in industrial CPS, large-scale AI models such as LLM can assist in the early detection of fault tolerance, suspicious data traffic, and intruder detection in the network by deeply analyzing the large dataset. Moreover, it can alert the network administrator early to take effective measures to overcome the issue.

\subsubsection{Situation-Aware Resource Management}

Situational awareness is essential for managing complex and dynamic wireless networks. In Internet-of-Vehicle (IoV) networks, DTs can leverage awareness of vehicular mobility and the 3D RF environment to optimize resource allocation and maximize spectrum utilization \cite{Zoheb_1}. Similarly, situational awareness is crucial for supporting extended reality (XR) applications in NextG and beyond communications. For instance, in future Industrial IoT (IIoT) networks, production engineers may use head-mounted display (HMD) devices for real-time production monitoring via augmented reality (AR) and virtual reality (VR). The integration of LLMs with DTs enhances the identification and prediction of dynamic network conditions and resource demands, enabling adaptive RRM. In IoV networks, LLM-DTNet can analyze and predict traffic congestion and road conditions, enabling the RRM engine to allocate radio resources more efficiently. Similarly, in IIoT networks, by leveraging real-time production data, LLM-DTNet can optimize resource allocation to provide real-time assistance to engineers, helping them make informed decisions.

\subsection{Use Case: Non-Terrestrial Networks}

\subsubsection{Proactive Resource Management}
In NextG communications, multiple frequency bands, including FR1 (sub-7 GHz), FR2 (7-24 GHz), and FR3 (24-52 GHz), are expected to coexist to meet growing bandwidth demands. NTN platforms such as low-altitude platforms (LAPs), high-altitude platforms (HAPs), and low-earth orbit (LEO) satellites may dynamically access the spectrum to support diverse applications. Emerging services may require NTN-NTN and NTN-terrestrial coexistence, necessitating advanced spectrum sharing and interference management. Conventional spectrum-sharing mechanisms, based on worst-case assumptions, often result in underutilized spectrum. Spectrum utilization architecture can play an important role by dynamically adjusting spectrum access policies according to static and dynamic contexts \cite[Table 1]{Zoheb_3}. The proposed LLM-DTNet architecture plays a critical role in designing and validating context-aware spectrum-sharing policies. By leveraging LLMs, AI/GenAI tools, and raw spectrum measurements (e.g., I/Q samples and spectrograms), LLM-DTNet can analyze and predict contexts such as user traffic demand, spectrum availability, and band propagation characteristics. 
This context-aware approach enables adaptive spectrum-sharing policies that proactively mitigate risks from outdated or uninformed decisions. For example, in spectrum sharing with a moving radar, if LLM-DTNet predicts radar activity, it can preemptively blank radar-occupied subcarriers to ensure seamless coexistence without disrupting radar operations.

\subsubsection{Situation-Aware Resource Management}
The situational awareness provided by the LLM-DTNet architecture enables automatic identification and resolution of network faults, such as call drops or network outages. This intelligent troubleshooting and self-healing capability can reduce customer complaints and significantly enhance end-user satisfaction. This feature is particularly valuable for leveraging NTN to address emergency situations.  For example, during a high-traffic event at a stadium, the nearby fixed cellular infrastructure may struggle to handle the sudden surge in user demand. In such scenarios, LAPs (enabled by drones) can be deployed to meet the increased connectivity needs. The optimal placement of these LAPs depends on factors such as network load, user density, backhaul capacity, and LAP energy constraints, all of which are critical for minimizing deployment costs while maximizing the users' satisfaction. LLM-DTNet, running on a nearby edge server, can efficiently solve this optimization problem, thanks to its real-time situational awareness.

\section{Future Research Directions}
The integration of LLMs and DTs in wireless networks presents several challenges, including communications efficiency, efficient multimodal data processing, and security concerns.  
Along this line, the following future research directions are proposed for LLM-DTNet. 

\noindent\textbf{\textit{Direction 1:} Handling Large Datasets for Training and Energy Efficient Design:} Handling a large dataset for training in the proposed LLM-DTNet framework is a big challenge. Developing task-specific LLM-DT integration, designing a federated LLM-DT approach, and prioritizing tasks can help address this issue. These strategies require thorough investigation and innovative implementation. The computational overhead of LLMs is substantial, leading to high energy consumption and requiring powerful hardware. 
A hierarchical model design incorporating lightweight edge models, adaptive architectures that dynamically adjust complexity based on context, custom low-power AI accelerators, and on-device caching can effectively mitigate computational challenges for LLMs.

\noindent\textbf{\textit{Direction 2:} Addressing Data Collection and Computation Cost of DT:} 
Constructing a trustworthy, multi-modal DT of wireless networks requires extensive data collection and significant computational resources, posing challenges for resource-limited environments. Overcoming these limitations involves two key steps: (i) employing data mining to identify context-critical parameters (both static and dynamic) from real or historical data, and (ii) formulating context acquisition as a priority-aware resource scheduling problem to capture rapidly changing contexts and mitigate synchronization gaps. Moreover, GenAI and graph neural network (GNN)-based AI models can fill data voids by predicting missing contexts, thereby strengthening the overall fidelity of DT emulation.

Furthermore, data collection and computation costs increase significantly for large-scale DTs, such as city-scale DT implementations. To mitigate these challenges, the Internet-of-Federated DTs paradigm aggregates multiple local DTs via federated learning \cite{Yu2024}, enabling privacy-preserving performance comparisons across different environments. Each local DT focuses on a small region, thereby reducing data collection and processing overhead. Nevertheless, optimizing the placement and updating of these local DTs remains essential to balance modeling accuracy with cost efficiency.

\noindent\textbf{\textit{Direction 3:} Secure and Privacy Preserving Model:} In the proposed framework, sensitive network and user data continuously flow between the physical network and its DT, creating potential attack surfaces for adversaries. These threats range from injecting malicious data to compromise the LLM’s decisions, eavesdropping on twin updates, or even manipulating the twin to feed false insights to network controllers. Developing a multi-layered security framework for LLM-DT integration in NextG—incorporating strong encryption, authentication mechanisms, and adversarial robustness techniques—can effectively safeguard data exchanges, mitigate prompt injections, and prevent unauthorized access. Privacy-preserving AI methods, such as federated learning, are promising to enable decentralized learning while minimizing data exposure, ensuring secure, adaptive, and resilient LLM-driven decision-making in next-generation wireless networks.

\section{Conclusions}
This article has proposed the LLM-DTNet architecture, which generates high-fidelity contextual data by leveraging a detailed and accurate DT model at each network layer. The data is then fine-tuned using task-specific AI/GenAI models at lower layers and LLMs at upper layers, enabling seamless network management for hierarchical resource allocation. The envisioned fusion of LLMs and DTs represents a potentially transformative paradigm for intelligent, resilient, and personalized communication networks, enhancing efficiency, autonomy, and user experience. However, further innovations are needed to develop layer-specific DT models while minimizing data collection and computational costs, designing efficient LLM architectures, and ensuring secure and seamless integration between LLM and DT layers.

\balance
\bibliographystyle{IEEEtran}
\bibliography{references_list}


 \section*{Biographies}
 \vspace{-1.2cm}
 \begin{IEEEbiographynophoto} {Majumder Haider} [S] (majumder.haider@bison.howard.edu) is a Ph.D. candidate in Electrical Engineering at Howard University, Washington, DC, USA. His current research interests include digital twin, intelligent reconfigurable surfaces (IRS), massive MIMO, and Industrial IoT networks. 
\end{IEEEbiographynophoto}

\vspace{-1.2cm}
\begin{IEEEbiographynophoto} {Imtiaz Ahmed} [SM] (imtiaz.ahmed@howard.edu) is an Assistant Professor at the Department of Electrical Engineering and Computer Science, Howard University, Washington, DC, USA. He works in the areas of wireless communications, signal processing, and computer networks.
\end{IEEEbiographynophoto}

\vspace{-1.2cm}
\begin{IEEEbiographynophoto} {Md. Zoheb Hassan} [M] (md-zoheb.hassan@gel.ulaval.ca) is an Assistant Professor at Laval University, QC, Canada. His research interests are Digital Twin, Spectrum Management, and Radio Resource Optimization.
\end{IEEEbiographynophoto}

 \vspace{-1.2cm}
\begin{IEEEbiographynophoto} {Kamrul Hasan} [M] (mhasan1@tnstate.edu) is an Assistant Professor in the Department of Electrical and Computer Engineering at Tennessee State University (TSU), where he leads cutting-edge research at the nexus of artificial intelligence (AI), transformer-based machine learning (ML), cybersecurity for NextG networks, and privacy-preserving secure data sharing.
\end{IEEEbiographynophoto}

\vspace{-1.2cm}
\begin{IEEEbiographynophoto} {H. Vincent Poor} [F] (poor@princeton.edu) is the Michael Henry Strater University Professor at Princeton University, where his interests include information theory, machine learning and network science, and their applications in wireless networks, energy systems, and related fields. He is a member of the U.S. National Academy of Engineering and the U.S. National Academy of Sciences, and a foreign member of the Royal Society and other national and international academies. He received the IEEE Alexander Graham Bell Medal in 2017.
\end{IEEEbiographynophoto}

\end{document}